\documentclass[preprint,amsmath,amssymb,superscriptaddress, floatfix,nofootinbib]
{revtex4-1}
\usepackage{epsfig,subfigure}
\usepackage{epstopdf}
\usepackage[utf8]{inputenc}
\usepackage{graphicx}
\usepackage{amssymb}
\usepackage{amsxtra}
\usepackage{amsmath}
\usepackage{booktabs,multirow,tabularx}
\usepackage{slashed}
\usepackage{float}
\usepackage{placeins}
\usepackage{color}
\usepackage{hyperref}
\usepackage{soul}
\usepackage{braket}
\usepackage{physics}
\usepackage[absolute]{textpos}

\begin{document}
\begin{textblock}{5}(11.8,1)
KEK-TH-2559
\end{textblock}

\title{CP-violating top-Higgs coupling in SMEFT}

\author{Vernon Barger  }
\email{barger@pheno.wisc.edu}
\affiliation{Department of Physics, University of Wisconsin, Madison, WI 53706 USA}
\author{ Kaoru Hagiwara}
\email{kaoru.hagiwara@kek.jp }
\affiliation{KEK Theory Center and Sokendai, Tsukuba, Ibaraki 305-0801, Japan}
\author{ Ya-Juan Zheng}
\email{yjzheng@iwate-u.ac.jp}
\affiliation{Faculty of Education, Iwate University, Morioka, Iwate 020-8550, Japan}
%

%
\begin{abstract}
The total cross section of the process
 $\mu^- \mu^+ \to \nu_\mu \bar{\nu}_\mu t \bar{t} H$
  has strong dependence on the CP phase $\xi$ of the top Yukawa coupling, where the ratio of $\xi=\pi$ and $\xi = 0$ (SM) grows to 670 at $\sqrt{s}$ = 30 TeV, 3400 at 100 TeV.   
We study the cause of the strong energy dependence and identify its origin as the $(E/m_W^{})^2$ growth of the weak boson fusion sub-amplitudes, $W_L^- W_L^+ \to t \bar{t} H$, with the two $W$’s are longitudinally polarized.   
We repeat the study in the SMEFT framework where EW gauge invariance is manifest and find that the highest energy cross section is reduced to a quarter of the complex top Yukawa model result, with the same energy power.   
By applying the Goldstone boson (GB) equivalence theorem, we identify the origin of this strong energy growth of the SMEFT amplitudes as associated with the dimension-6 $\pi^- \pi^+ ttH$ vertex, where $\pi^\pm$ denotes the GB of $W^\pm$.  
We obtain the unitarity bound on the coefficient of the SMEFT operator by studying all $2\to2$ and $2\to3$ cross sections in the $J=0$ channel.
\end{abstract}

\maketitle

Possible CP violation in the largest coupling of the SM,
the top Higgs Yukawa coupling, has received interest
because of its potential role in producing the baryon
asymmetry of the universe~ \cite{Zhang:1994fb,Fuchs:2020uoc,Bahl:2022yrs}.
The coupling has been measured at the LHC in $t\bar{t}H$ production\cite{ATLAS:2018mme,CMS:2020cga,ATLAS:2020ior,ATLAS:2023cbt}, and in single top plus Higgs
production~\cite{CMS:2018jeh,ATLAS:2020ior,ATLAS:2023cbt}. 
Most phenomenological studies of CP asymmetries in
the above processes have been performed by adopting a complex Yukawa coupling, which can be parametrized
as
\begin{eqnarray}
{\cal L}_{ttH}=-g H\bar{t}(\cos\xi+i\gamma_5\sin\xi)t
\label{eq:CYlag}
\end{eqnarray}
with real positive $g$ and $|\xi|\leq\pi$.
The SM Yukawa coupling is recovered by setting
$
g = g_{\rm SM} = m_t/v
$
where $v=246$ GeV is the SM Higgs VEV. 
The above parametrization, or its variants such as
$g\cos\xi = g_{\rm SM} \kappa_H$ (CP-even), $g \sin\xi = g_{\rm SM} \kappa_A$ (CP-odd), 
have been adopted in studying CP violating asymmetries
which change sign according to the sign of CP phase $\xi$,
in $pp \to t\bar{t}H$~\cite{He:2014xla,Bahl:2020wee,Bahl:2021dnc,Martini:2021uey}, $e^-e^+ \to t\bar{t}H$~\cite{Zhang:1994fb,Hagiwara:2017ban,Cheung:2023qnj}, single top plus $H$
production at the LHC~\cite{Barger:2018tqn,Barger:2019ccj,Bahl:2021dnc,Martini:2021uey}.

\begin{figure}[b]
\begin{center}
\subfigure[]{\includegraphics[width=0.45\textwidth,clip]{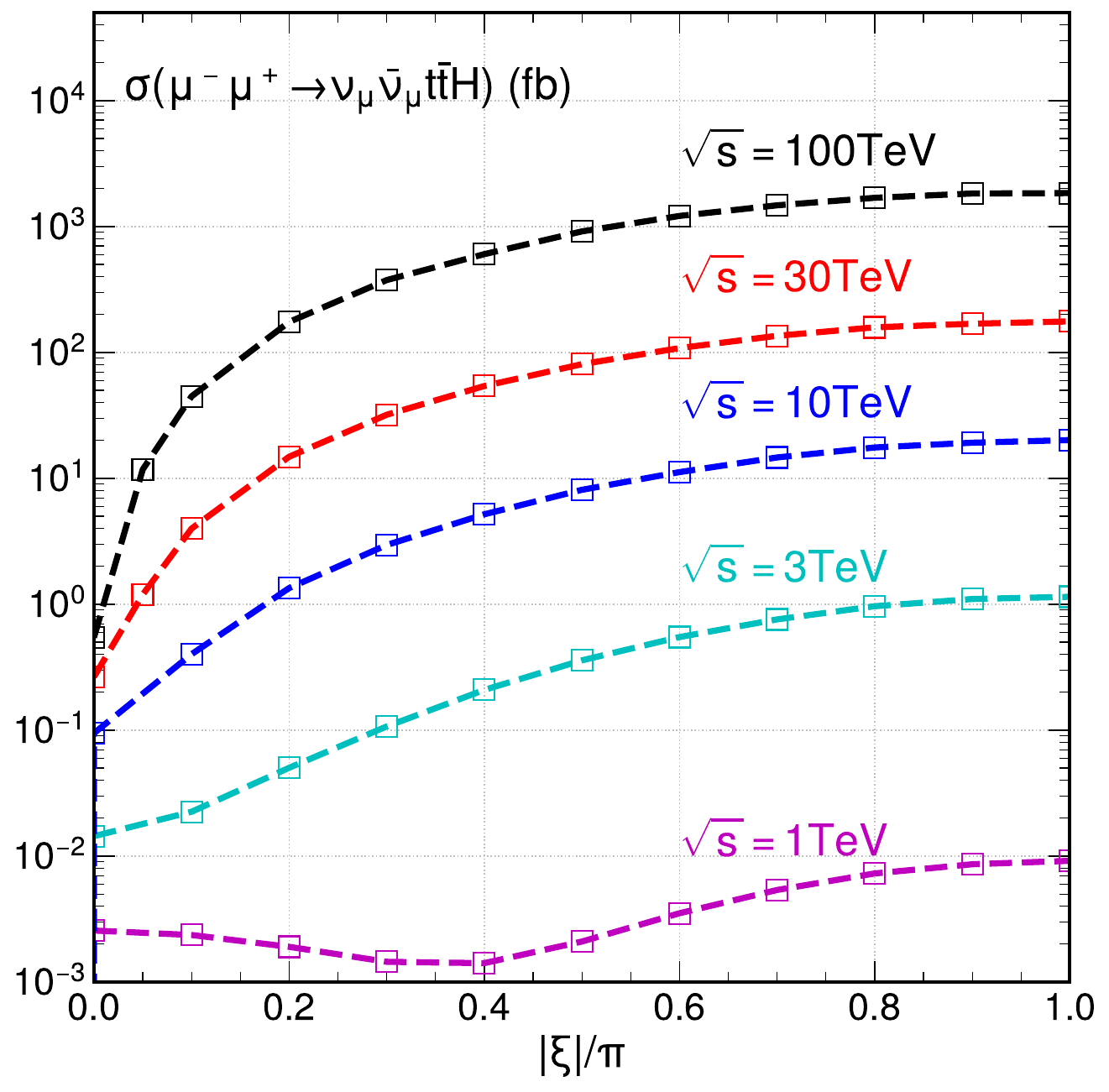}}
\subfigure[]{\includegraphics[width=0.45\textwidth,clip]{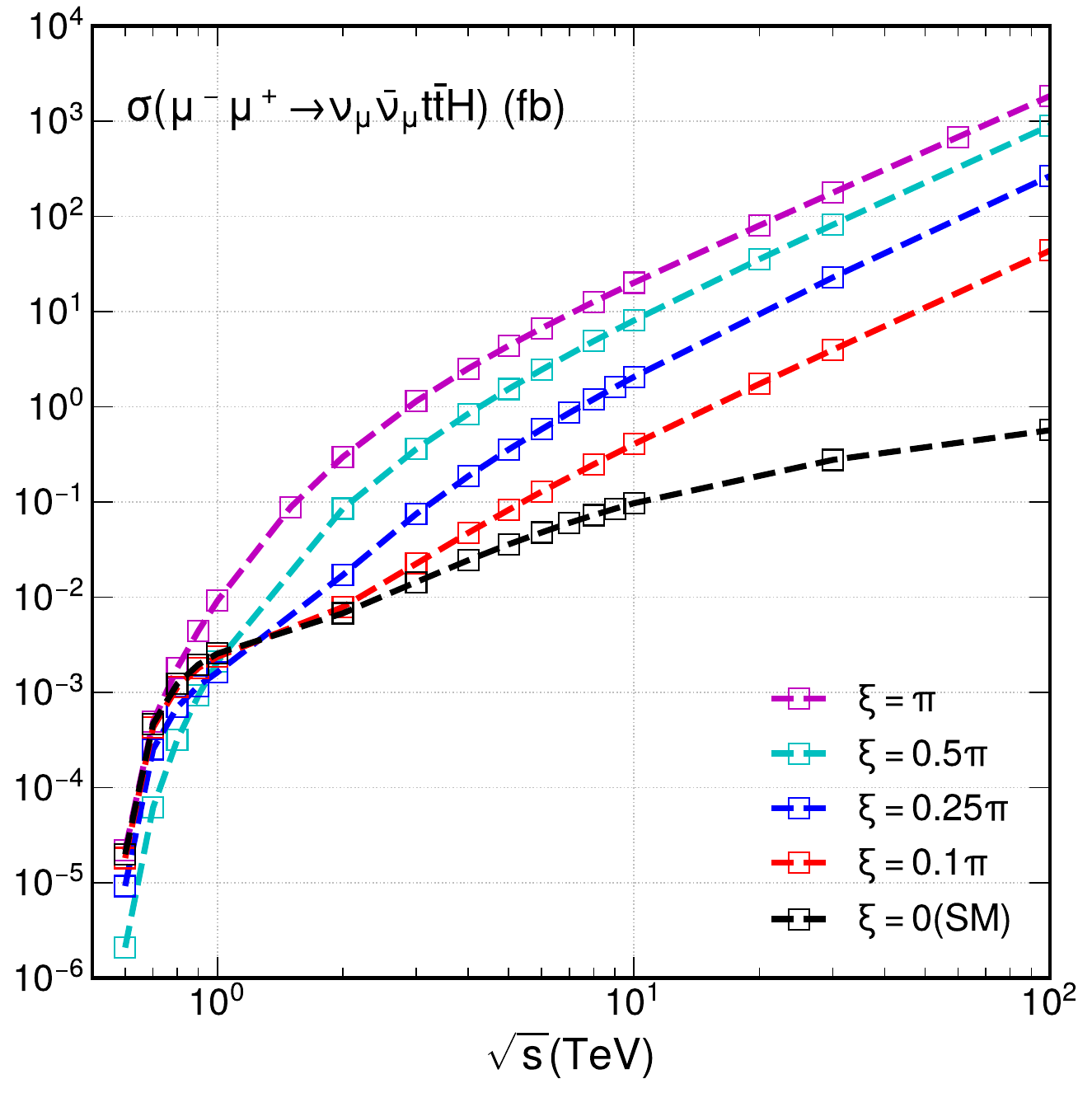}}
\end{center}
\caption{Total cross section of $\nu_\mu\bar{\nu}_\mu t\bar{t}H$ production at a muon collider: (a) $\xi$ dependence at several energies (b)  $\sqrt{s}$ dependence at several $\xi$ values.}
\label{fig:tthvv_CYxs}
\end{figure}
In this paper, we report our findings on the use of
the above CP violating Yukawa coupling, at future
high energy colliders, in particular at muon colliders ~\cite{AlAli:2021let,Aime:2022flm}. 
Shown in Fig.\,\ref{fig:tthvv_CYxs} is the total cross section of the process
\begin{eqnarray}
\mu^-\mu^+\to \nu_\mu\bar{\nu}_\mu t\bar{t}H
\label{proc:mmttH}
\end{eqnarray}
as a function of $ \xi$ for colliding muon energies in the range 0.6 TeV $\leq \sqrt{s} \leq$ 100 TeV.
In the left hand side, Fig.\,\ref{fig:tthvv_CYxs}(a), we show the cross section vs.\,$|\xi|/\pi$ at $\sqrt{s} =1,3,10,30,100$ TeV, whereas in the right hand
side, Fig.\,\ref{fig:tthvv_CYxs}(b), the $\sqrt{s}$ dependence of the cross section is
shown for $|\xi|$ = 0 (SM), $0.1\pi$, $0.25\pi$, $0.5\pi$ and $\pi$.

In Fig.\,\ref{fig:tthvv_CYxs}(a), we note a sharp rise of the cross section
between $\xi=0$ (SM) and $|\xi|\simeq0.1\pi$ at very high energies
$(\gtrsim10$ TeV), whereas in Fig.\,\ref{fig:tthvv_CYxs}(b), we identify the
quadratic energy behavior $(\sqrt{s})^2 = s$ of the total cross
section above $\sqrt{s} \simeq 10$ TeV, for all the non-zero $\xi$ cases.
In contrast, the SM cross section grows only logarithmically. 
We should understand the cause of this power behavior,
and the importance on phenomenological studies on CP asymmetries in the process Eq.\,(\ref{proc:mmttH}).

\begin{figure}[t]
{\includegraphics[scale=0.25]{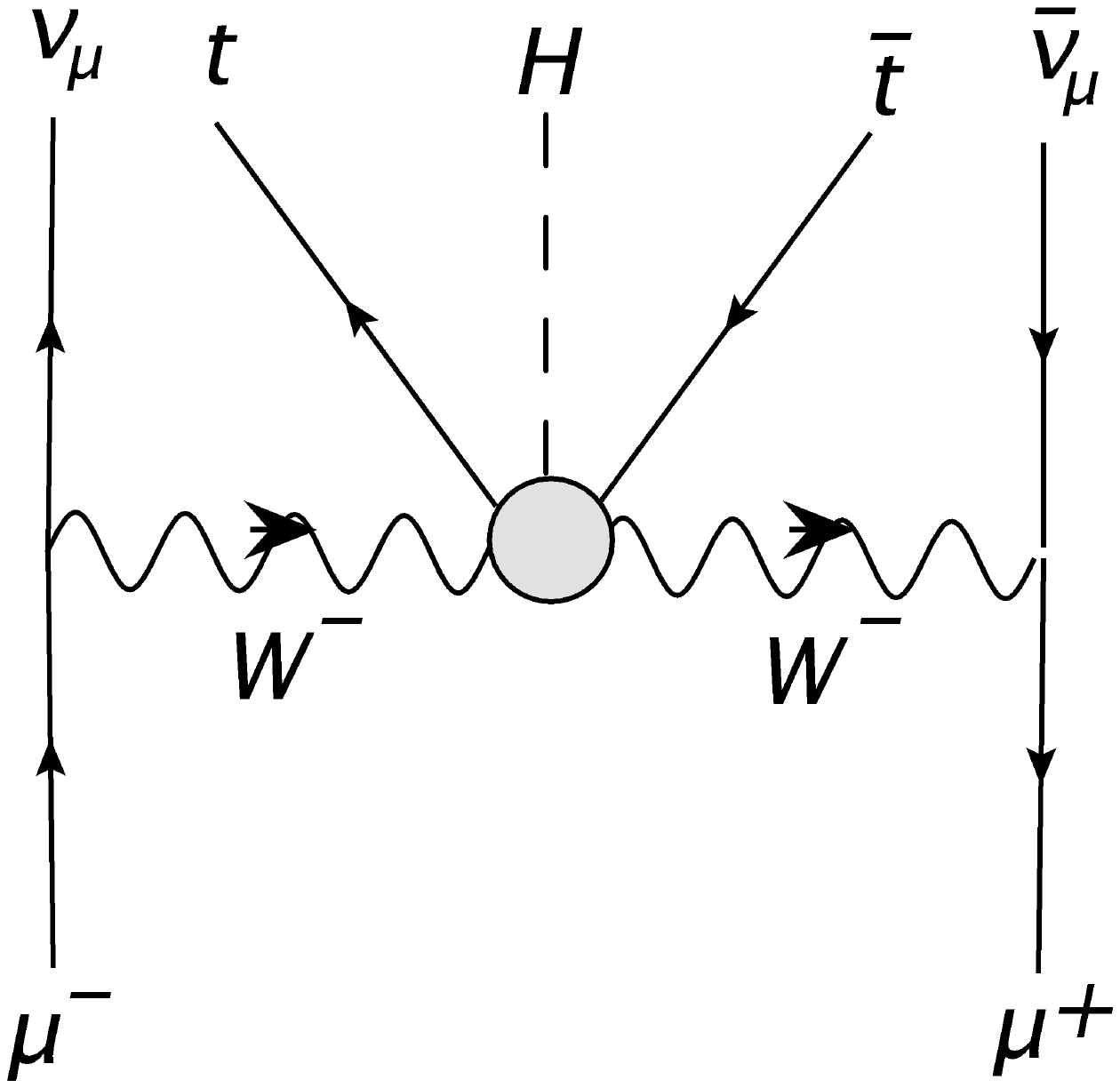}}
\caption{Weak boson fusion subdiagrams, contributing to the process
$\mu^- \mu^+ \to \nu_{\mu}\bar{\nu}_{\mu} t\bar{t} H$. }
\label{fig:d4feyn}
\end{figure}

We study all the Feynman diagrams generated by  {\tt MadGraph5\_aMC@NLO}\cite{Alwall:2014hca},
and identify that among 88 diagrams, 20 diagrams can
be categorized as weak boson fusion (WBF) subprocesses,
as depicted in Fig.\,\ref{fig:d4feyn}.
Since their contribution can be evaluated by making use
of the weak boson PDF, we calculate the total cross
section for the process
\begin{eqnarray}
W^-(q,h=0) ~W^+(\bar{q},\bar{h}=0) \to t \bar{t} H
\label{proc:wwhtt}
\end{eqnarray}
where helicities $h=\bar{h}=0$ represent longitudinal polarizations.
\begin{figure}[t]
\subfigure[]
{\includegraphics[width=0.45\textwidth,clip]{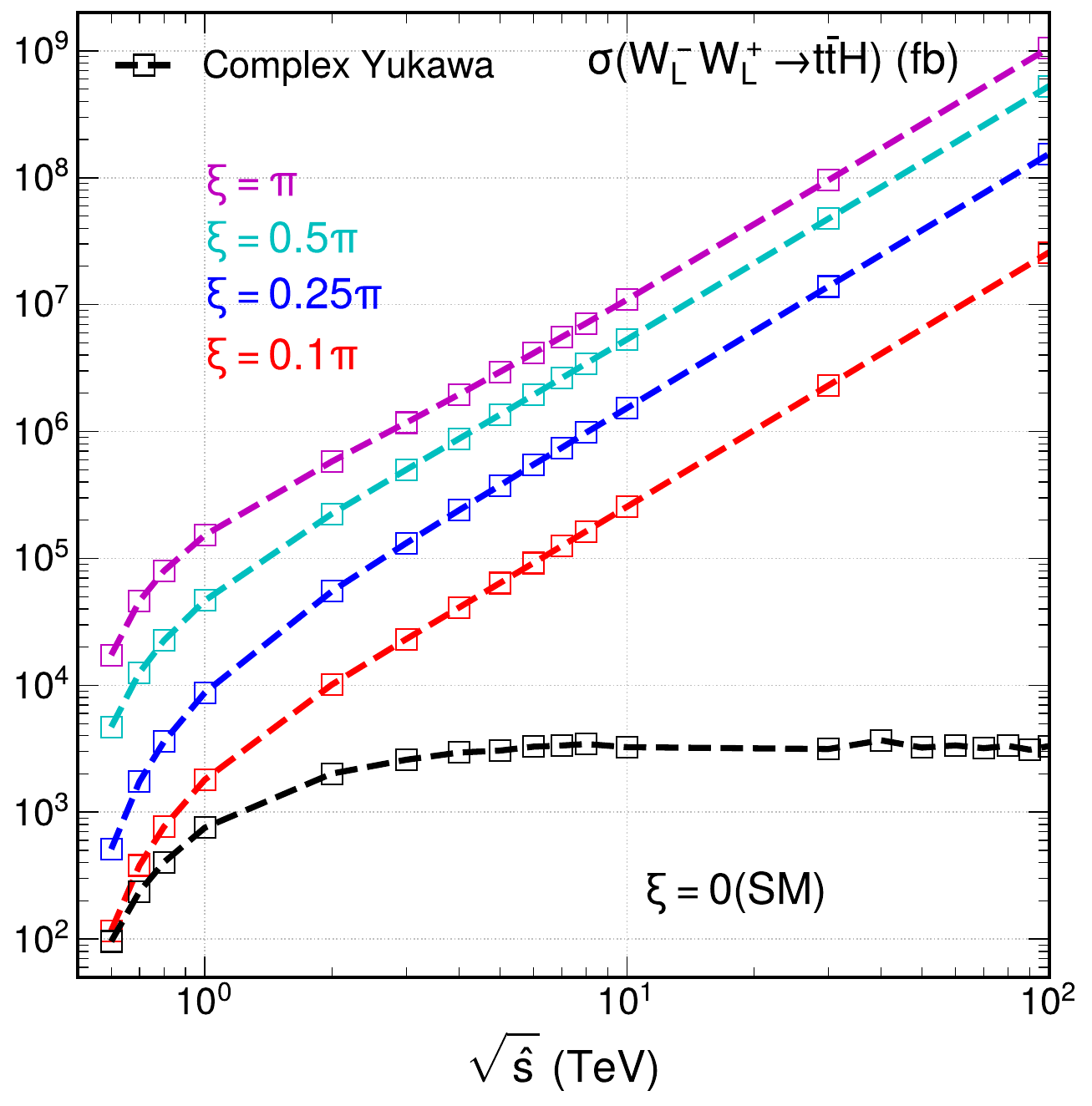}}
\subfigure[]
{\includegraphics[width=0.45\textwidth,clip]{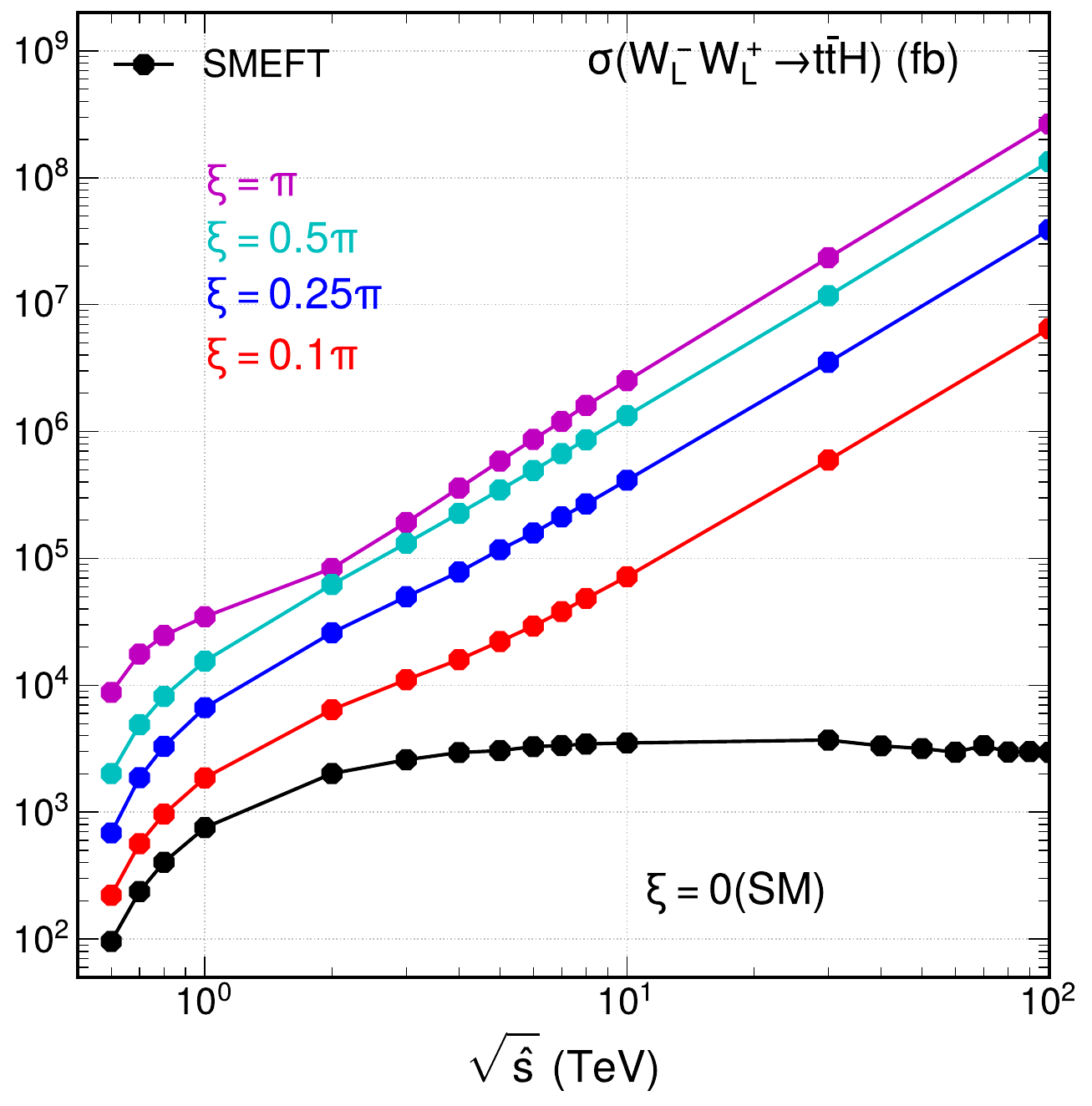}}
\caption{ $W_L^-W_L^+\to t\bar{t}H$  cross section vs.\,the colliding $W^-W^+$ energy $\sqrt{\hat{s}}$: (a) complex Yukawa model and (b) SMEFT.}
\label{fig:WWttH}
\end{figure}

Shown in Fig.\ref{fig:WWttH}(a) is the total cross section as
a function of $\sqrt{\hat{s}}$, the colliding $W^-W^+$ or the
$t\bar{t}H$ invariant mass, between
0.6 and 100 TeV. 
We can clearly identify the quadratic energy behavior
of the total cross section for all non-zero $\xi$
cases ($0.1\pi \leq |\xi| \leq \pi$) at energies above a few TeV.
The quadratic energy power behavior of the total cross
section is due to two powers of energy in the amplitudes from the incoming longitudinally
polarized weak boson wave functions, which behave as
$E/m_W$ with $E= \sqrt{\hat{s}}/2$. 
In the SM, such powers of $E/m_W$ are present in
individual Feynman amplitudes, but they cancel 
after summing over all Feynman amplitudes,
leading to the Goldstone boson equivalence theorem
(GBET) as a manifestation of gauge invariance~\cite{Cornwall:1974km,Chanowitz:1985hj}.

We therefore look for gauge invariant formulation
with a CP violating Yukawa coupling, such as arises
from  a two Higgs doublet model with a CP violating
Higgs potential.
When all the non-SM degrees of freedom are heavy,
all such models can be reduced to SMEFT~\cite{Leung:1984ni,Buchmuller:1985jz,Grzadkowski:2010es,Maltoni:2019aot}, and we
identify the following top quark Yukawa sector\cite{Zhang:1994fb,Whisnant:1994fh}
\footnote{The use of the SMEFT operator in Eq.\,(\ref{eq:op})
has been suggested by Cen Zhang to one of us in
the year 2020, before he tragically passed away
in 2021.
The dimension-6 operator in eq.(\ref{eq:op}) is named $Q_{u\phi}$ in the Warsaw basis~\cite{Grzadkowski:2010es}.}
\begin{eqnarray}
{\cal L}&=&-y_{\rm SM}Q^\dagger\phi t_R
+\frac{\lambda}{\Lambda^2}Q^\dagger\phi t_R\phi^\dagger\phi
+{\rm h.c.},
\label{eq:op}
\end{eqnarray}
with $\lambda$ a complex number denoting the deviation from the SM. 
By inserting the component fields $Q = (t_L, b_L)^T$ and
$\phi = ( (v+H+i\pi^0)/\sqrt{2}, i\pi^- )^T$ into the Lagrangian Eq.\,(\ref{eq:op}),
where $\pi^0$ and $\pi^\pm$ are the Goldstone bosons of
$Z$ and $W^\pm$, respectively, 
we find
\begin{eqnarray}
{\cal L}_{ttH}^{ \rm SMEFT}
  =
-Q^\dagger\phi t_R\left[y'-\frac{\lambda}{\Lambda^2}
\left(vH+\frac{H^2+(\pi^0)^2}{2}+\pi^+\pi^-\right)\right] +{\rm h.c.},
\label{eq:smeft}
\end{eqnarray}
where
%
$\displaystyle{y' = y_{\rm SM} - \frac{\lambda v^2}{2\Lambda^2}}$
%
is the Yukawa coupling including the SMEFT operator
contribution\footnote{When the original Yukawa coupling $y$ and
the coefficient $\lambda$ of the dimemsion-6 operator have
the same flavor structure, we arrive at the flavor
diagonal vertices as above.  For a general treatment,
see e.g. ref.\cite{Brod:2022bww}.}.
With $\displaystyle{m_t = \frac{|y^\prime|}{\sqrt{2}} v}$, the phase $\arg(y^\prime)$ is absorbed by $t_R$ and $\lambda$ is re-phased accordingly.
In the basis where we denote $t_R$ and $\lambda$ after the re-phasing, 
we can express the SMEFT Lagrangian  Eq.\,(\ref{eq:smeft}) as
\begin{eqnarray}
{\cal L}_{ttH}^{\rm SMEFT} =
 &&
  -m_t t_L^\dagger t_R
  -g_{\rm SM} \left[ (H+i\pi^0)t_L^\dagger+i\sqrt{2}\pi^- b_L^\dagger \right] t_R
  \nonumber
  \\
 &&
  +(g_{\rm SM}-g e^{i\xi}) \left\{ H t_L^\dagger t_R
  +\frac{H}{v}\left[ (H+i\pi^0)t_L^\dagger+i\sqrt{2}\pi^- b_L^\dagger\right]t_R \right\}
    \nonumber
  \\
 &&
  +(g_{\rm SM}-g e^{i\xi}) \Bigg\{
  \left[\frac{H^2+(\pi^0)^2}{2v}+\frac{\pi^+\pi^-}{v}\right] t_L^\dagger t_R
      \nonumber
  \\
 &&
  +\frac{H^2+(\pi^0)^2+2\pi^+\pi^-}{2v^2}
  \left[(H+i\pi^0)t_L^\dagger +i\sqrt{2}\pi^-b_L^\dagger \right] t_R \Bigg\}
  +h.c.,
  \label{eq:SMEFTlag}
\end{eqnarray}
where 
\begin{eqnarray}
{g_{\rm SM} = \frac{m_t}{v}}
\quad
\quad
{\rm and}
\quad
\quad
{ g_{\rm SM}-g e^{i\xi} = \frac{\lambda v^2}{\sqrt{2}\Lambda^2}}.
\label{eq:geixi-gsm}
\end{eqnarray}
We maintain the original gauge invariant structure
of the couplings in the above expression which agrees with ref.\cite{Brod:2022bww}.  When we drop
all terms proportional to Goldstone boson fields,
it reduces to the results in ref.\cite{Whisnant:1994fh}.

A few comments are in order.
First, the $ttH$ coupling of the SM, $g_{\rm SM}$, is replaced by the
complex coupling $g e^{i\xi}$, which is identical to the
phenomenological Lagrangian Eq.\,(\ref{eq:CYlag}). 
Second, although the $ttH$ coupling is changed, the dimension-4
part of the Goldstone boson couplings remain the same
as the SM.
Third, the $ttHH$ coupling appears in the third and fourth terms,
whereas the $ttHHH$ coupling appears in the last term.
Fourth, all the vertices in the last term of the above
Lagrangian have mass dimension-6, with 3 boson fields
and a pair of fermion fields.

\begin{figure}[t]
\subfigure[]
{\includegraphics[width=0.26\textwidth,clip]{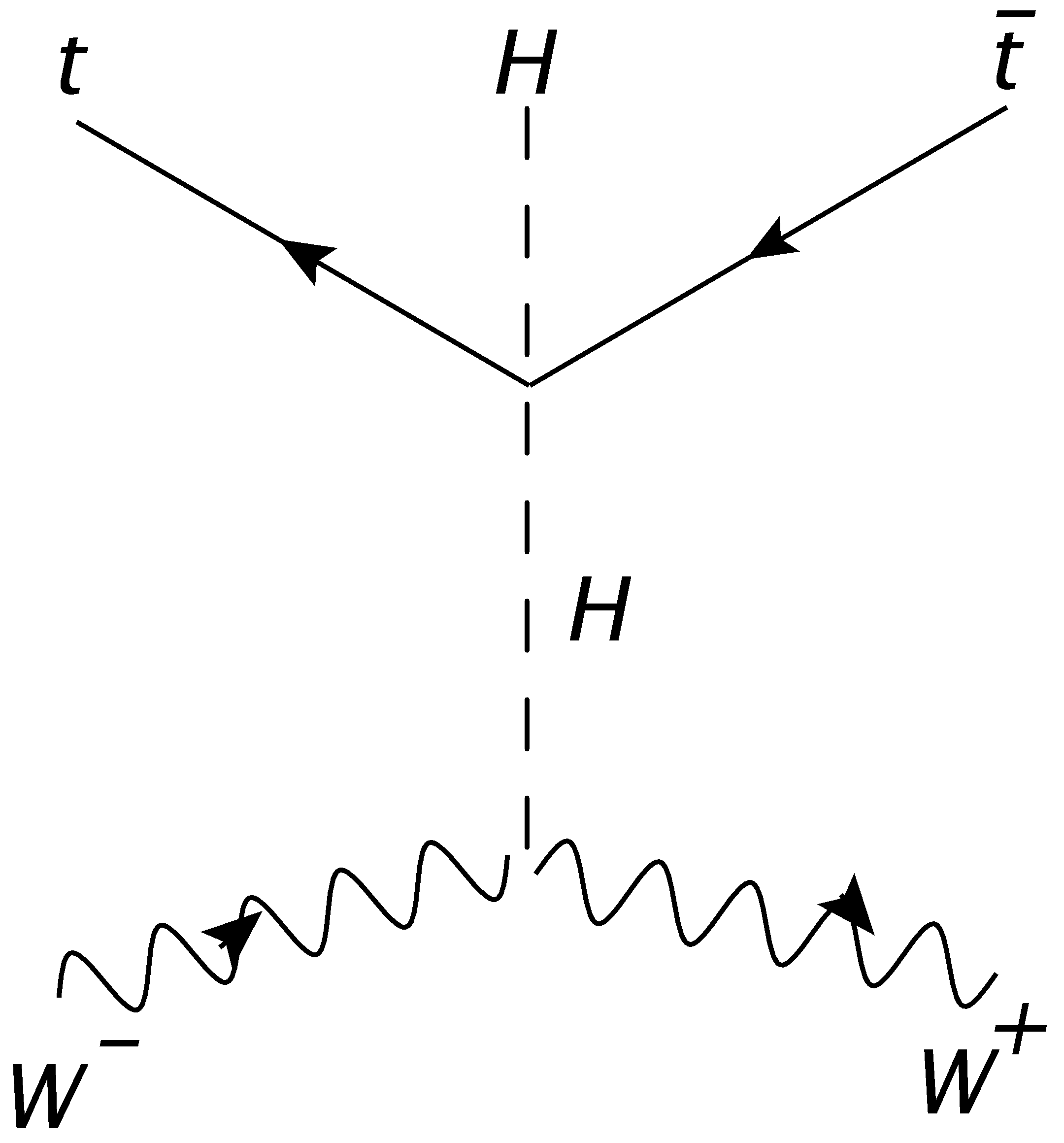}}
\hfill%
\subfigure[]
{\includegraphics[width=0.29\textwidth,clip]{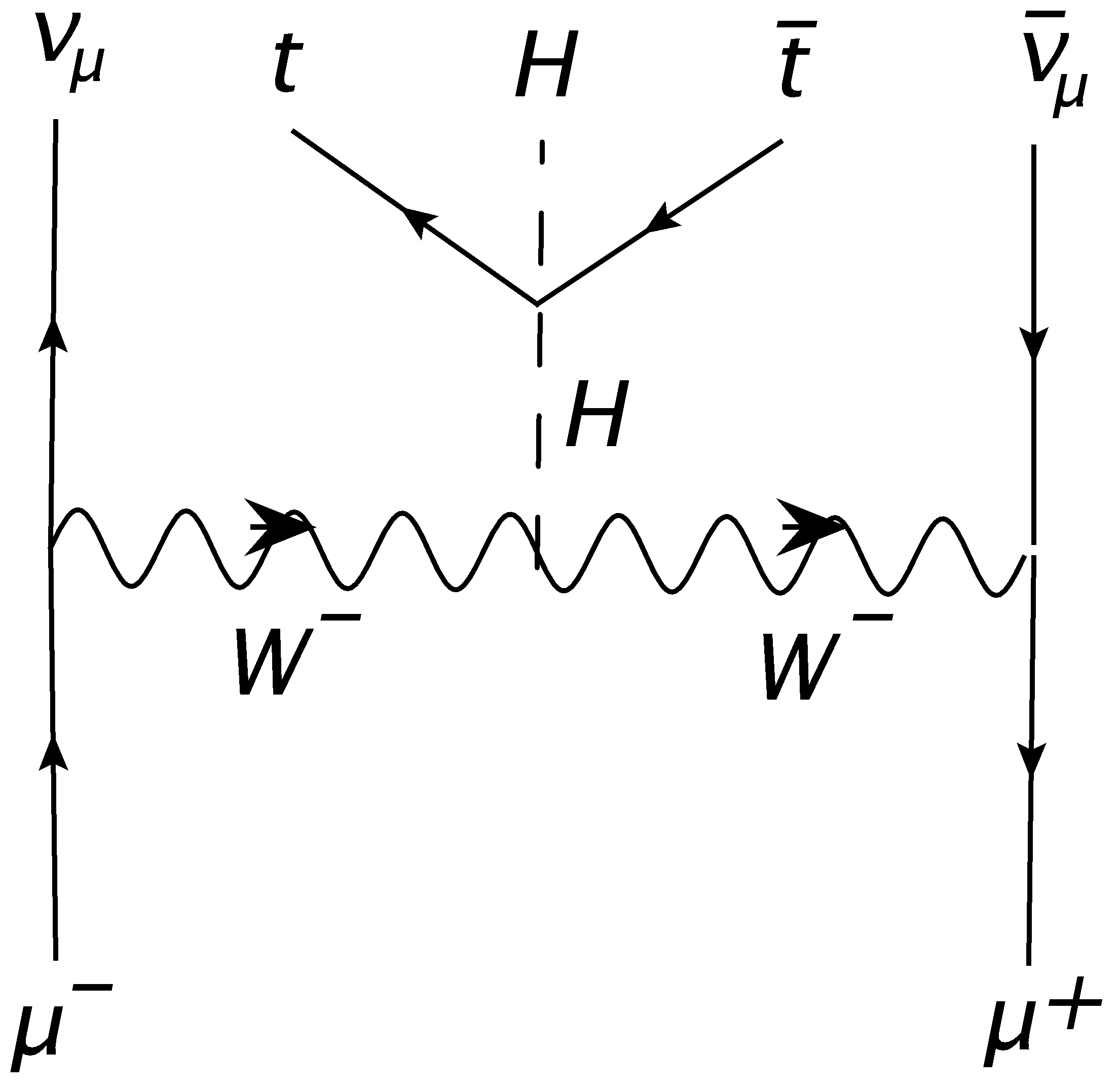}}
\hfill%
\subfigure[]
{\includegraphics[width=0.28\textwidth,clip]{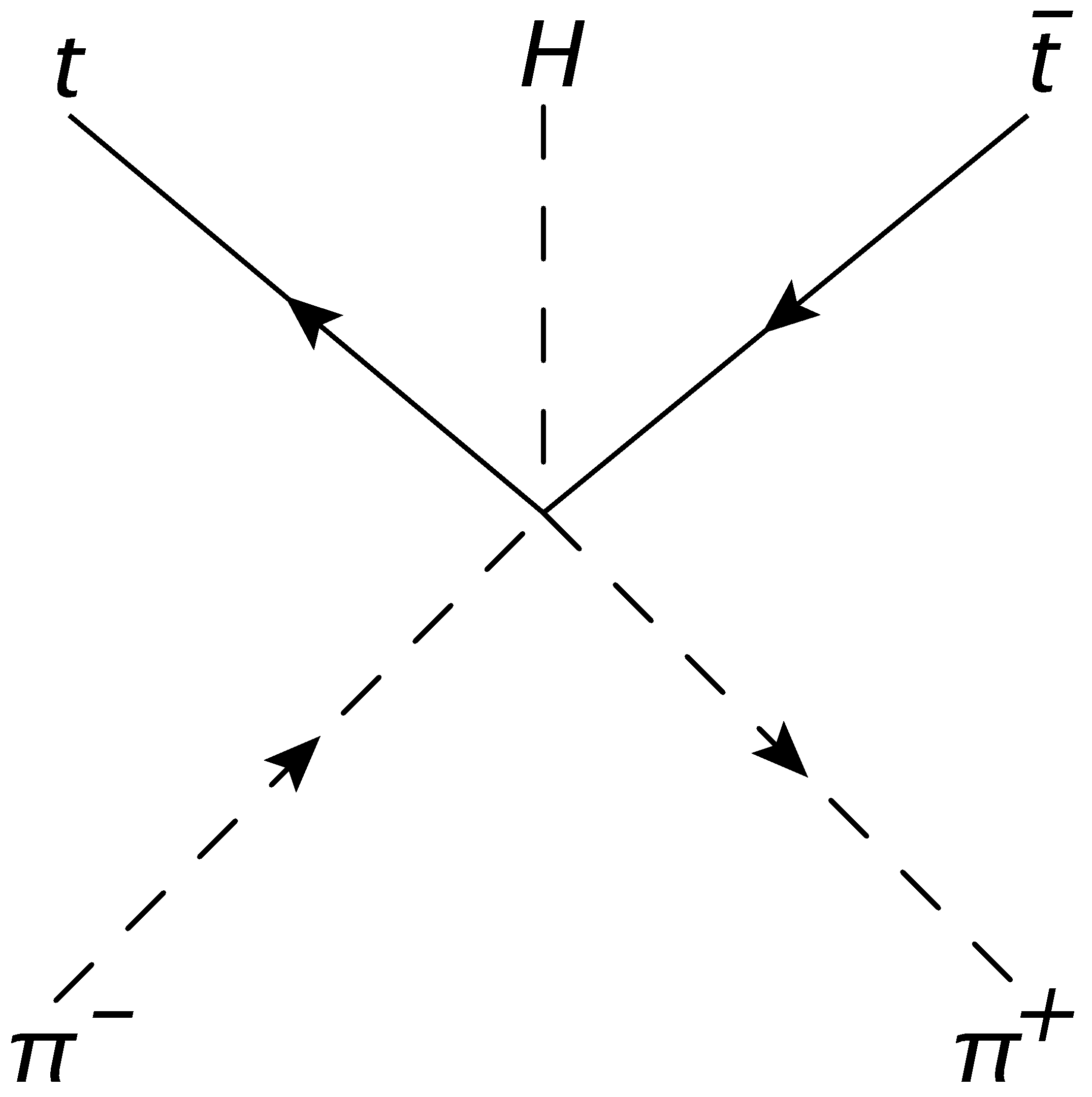}}
\caption{ Additional Feynman diagrams  in SMEFT contributing to the process
(a) $W^-_LW^+_L\to  t\bar{t}H$, 
(b) $\mu^-\mu^+\to \nu_\mu\bar{\nu}_\mu t\bar{t}H$ in the unitary gauge, and 
(c) 5-point contact vertex $\pi^-\pi^+ttH$, giving the high energy limit of the $W^-_LW^+_L\to  t\bar{t}H$ amplitudes. }
\label{fig:feynman3}
\end{figure}

In the unitary gauge, only the $ttHH$ coupling 
\begin{eqnarray}
{\cal L}_{ ttHH}^{\rm SMEFT}
  =
\frac{3(g_{\rm SM}-ge^{i\xi})}{v}\frac{H^2}{2}t_L^\dagger t_R + {\rm h.c.},
\end{eqnarray}
contributes
to the weak boson fusion process Eq.\,(\ref{proc:wwhtt}) and to the muon
collider process of Eq.\,(\ref{proc:mmttH}).
We modify the {\tt HELAS} code\cite{Hagiwara:1990dw,Murayama:1992gi} of {\tt MadGraph5} to evaluate the amplitude of the Feynman diagram in Fig.\,\ref{fig:feynman3}(a) in the $WW$ fusion process Eq.\,(\ref{proc:wwhtt}) and also the diagram Fig.\,\ref{fig:feynman3}(b) in the muon collider process Eq.\,(\ref{proc:mmttH}).

By adding the diagram Fig.\,\ref{fig:feynman3}(a) to the 20 diagrams generated
by {\tt MadGraph5} with the CP violating Yukawa coupling, we obtain the SMEFT amplitudes
\begin{eqnarray}
{\cal M}(W_L^-W_L^+\to t\bar{t}H)_{\rm SMEFT}
=\sum_{k=1}^{20}{\cal M}_k+{\cal M}_{\rm Fig.\,\ref{fig:feynman3}(a)},
\label{eq:wwtth21}
\end{eqnarray}
which gives the total cross section in Fig.\ref{fig:WWttH}(b) of the weak boson fusion process Eq.\,(\ref{proc:wwhtt}).
We notice a significant reduction of the cross section at
all energies, as compared to the results of the complex
Yukawa coupling model which are shown in Fig.\ref{fig:WWttH}(a).
On the other hand, we find that the high energy power
behavior of the total cross section is the same as
that of the complex Yukawa model.
Cross comparison of the two results, we find that 
when
$\sqrt{\hat{s}}\gtrsim10~{\rm TeV}$
\begin{eqnarray}
\sigma_{\rm tot}(W_L^-W_L^+ \to ttH)_{\rm SMEFT}
  \approx
\frac{1}{4} \sigma_{\rm tot}(W_L^-W_L^+ \to ttH)_{\rm complex~Yukawa}
\end{eqnarray}
for the same value of the complex Yukawa coupling,
$g e^{i\xi}$.

In order to clarify our findings, we calculate the
Goldstone boson scattering amplitudes for the process
$\pi^- \pi^+ \to t \bar{t} H$
analytically, by using the SMEFT Lagrangian Eq.\,(\ref{eq:SMEFTlag}). 
At high energies, the dimension-6 $\pi^-\pi^+ ttH$ vertex
contribution dominates, as depicted by the diagram Fig.\,\ref{fig:feynman3}(c), and we find
\begin{eqnarray}
{\cal M}^{\pm\pm}_{\rm Fig.\,\ref{fig:feynman3}(c)}
  =
\frac{1}{v^2}\left[\mp2p_t \left(g_{\rm SM}-g\cos\xi\right) -i m_{tt}\left(g\sin\xi\right)\right]
\label{eq:pipitth}
\end{eqnarray}
where $p_t$ is the magnitude of $t$ and $\bar{t}$ momentum in the $t\bar{t}$ rest frame, $\pm\pm$ denotes $t$ and $\bar{t}$ helicities
in the same frame, which should be common.
Although the Higgs boson energy does not appear in the
amplitude in Eq.\,(\ref{eq:pipitth}), it is fixed as
$
E_H = \displaystyle{\frac{\sqrt{\hat{s}}}{2} \left(1+\frac{m_H^2-m_{tt}^2}{\hat{s}}\right)},
$
in the colliding $W^-W^+$ or $t\bar{t}H$ rest frame.
Because the amplitude in Eq.\,(\ref{eq:pipitth}) grows with the invariant
mass of the $t\bar{t}$ pair, $m_{tt}$, soft Higgs boson with energetic
$t$ and $\bar{t}$ configuration dominates the total cross section
at high energies. 
The total cross section
\begin{eqnarray}
\sigma_{\rm tot}(\pi^-\pi^+\to t\bar{t}H)
  =
\frac{1}{2\hat{s}} \sum_{h,\bar{h}=\pm 1/2}
\int |{\cal M}^{h\bar{h}}|^2 d\Phi_{t\bar{t}H},
\end{eqnarray}
obtained by using the above analytic amplitudes agree
with the total cross section obtained by the numerical
code in the unitary gauge at high energies.
The linearly rising curves with the quadratic power of
$\sqrt{\hat{s}}$ are reproduced by the Goldstone boson
scattering cross section, verifying the GBET.

We therefore confirm that the SMEFT realization of
the complex Yukawa coupling model\cite{Zhang:1994fb,Whisnant:1994fh} reproduces all
low energy phenomenology of the processes which
are not affected by the $ttHH$ coupling,
and gives cross sections which are consistent
with the GBET, as a consequence of the gauge invariance.
However, the mystery remains.
Why is the total cross section of the complex Yukawa model at high energies in Fig.\ref{fig:WWttH}(a) four times the SMEFT cross section in  Fig.\ref{fig:WWttH}(b)?

In an attempt to clarify the above mystery, we
compute analytically the only one additional diagram of the
SMEFT in unitary gauge, Fig.\ref{fig:feynman3}(a).
We find that the only non-vanishing amplitudes are
\begin{eqnarray}
{\cal M}_{\rm Fig.\,\ref{fig:feynman3}(a)}^{\pm\pm}
  =
  \frac{3}{v^2}
  \left[ \mp 2p_t(g_{\rm SM}-g\cos\xi) -im_{tt}(g\sin\xi) \right]
    \frac{(\hat{s}-2m_W^2)}{(\hat{s}-m_H^2)},
\end{eqnarray}
where the $t$ and $\bar{t}$ have the same helicities,
$h = \bar{h} = \pm 1/2$, in the $t\bar{t}$ rest frame. 
Comparing the above amplitudes with the Goldstone
boson amplitudes Eq.\,(\ref{eq:pipitth}), we find
\begin{eqnarray}
{\cal M}_{\rm Fig.\,\ref{fig:feynman3}(a)}^{\pm\pm}
  = 
3 {\cal M}^{\pm\pm}_{\rm Fig.\,\ref{fig:feynman3}(c)}\cdot\left\{ 1 + {\cal O}(\frac{1}{\hat{s}}) \right\}.
\label{eq:fig4a3fig4c}
\end{eqnarray}
On the other hand, the GBET for ${\cal M}(W_L^-W_L^+\to t\bar{t}H)$ in Eq.\,(\ref{eq:wwtth21}) gives 
\begin{eqnarray}
\sum_{k=1}^{20} {\cal M}_k + {\cal M}_{\rm Fig.\,\ref{fig:feynman3}(a)} = {\cal M}_{\rm Fig.\,\ref{fig:feynman3}(c)}\cdot\left\{ 1 + {\cal O}(\frac{1}{\hat{s}}) \right\}.
\end{eqnarray}
Inserting Eq.\,(\ref{eq:fig4a3fig4c}), we find
\begin{eqnarray}
\sum_{k=1}^{20} {\cal M}_k \approx -2 {\cal M}_{\rm Fig.\,\ref{fig:feynman3}(c)},
\end{eqnarray}
at high energies.
This explains why the sum of all the diagrams with
CP violating Yukawa coupling gives 4 times the SMEFT
cross section at high energies.

It suggests that there exists a Higgs sector, which reproduces the complex Yukawa
coupling at dimension-4, but has no $ttHH$ coupling at dimension-5, and has the
contact $\pi^-\pi^+ ttH$ coupling at dimension-6 with minus two times that of the SMEFT Lagrangian Eq.(\ref{eq:SMEFTlag}).
Within the framework of SMEFT, such Higgs sector is indeed found at dimension-8 level
\footnote{We thank Kun-Feng Lyu for suggesting us to examine dimension-8 operators.}.
In this report, however, we proceed to study non-Standard Yukawa couplings in SMEFT at dimension-6 as in Eq.(\ref{eq:SMEFTlag}), which has no additional free parameters.
\begin{figure}[t]
\subfigure[]
{\includegraphics[width=0.4\textwidth,clip]{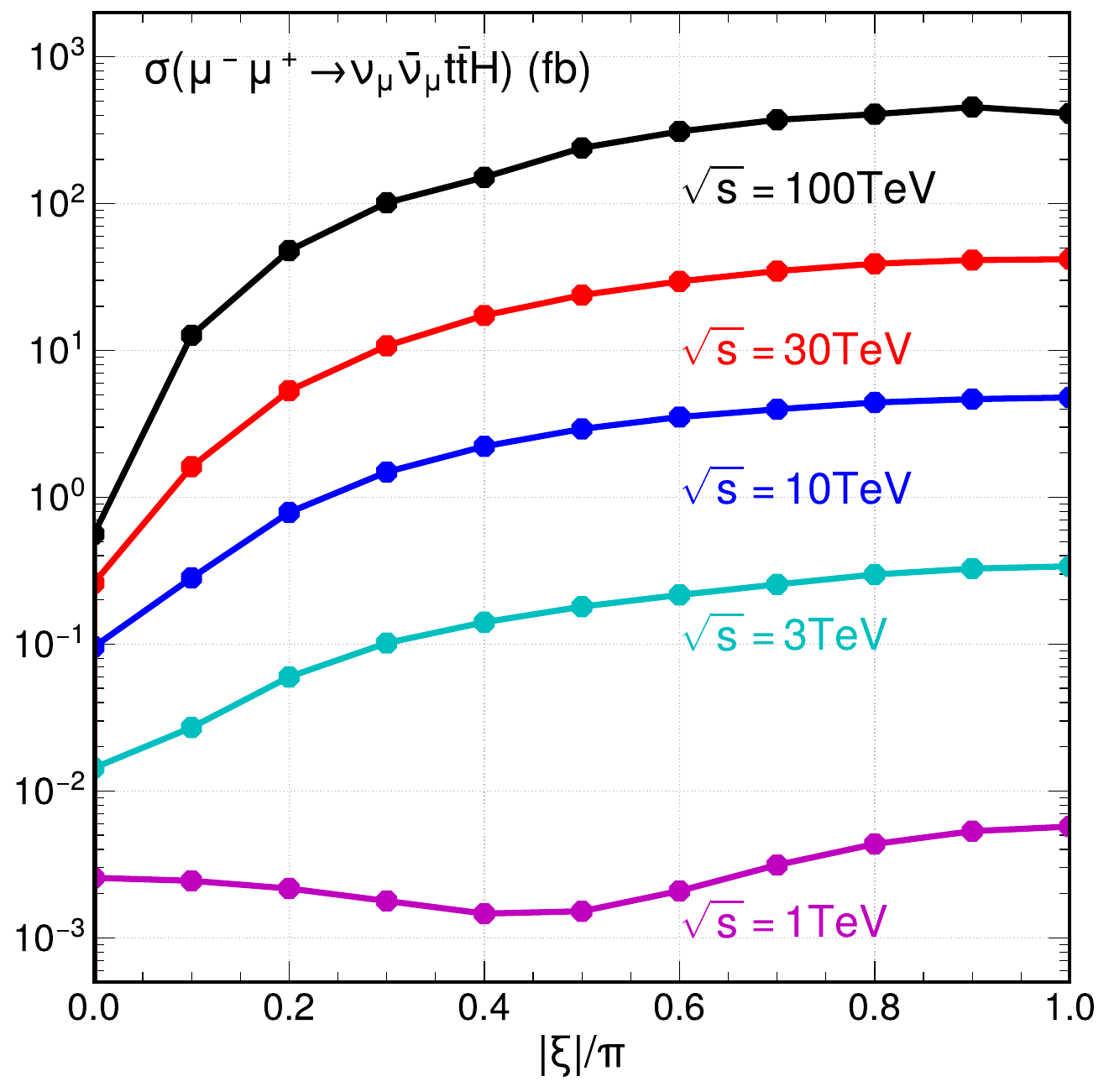}}
\subfigure[]
{\includegraphics[width=0.4\textwidth,clip]{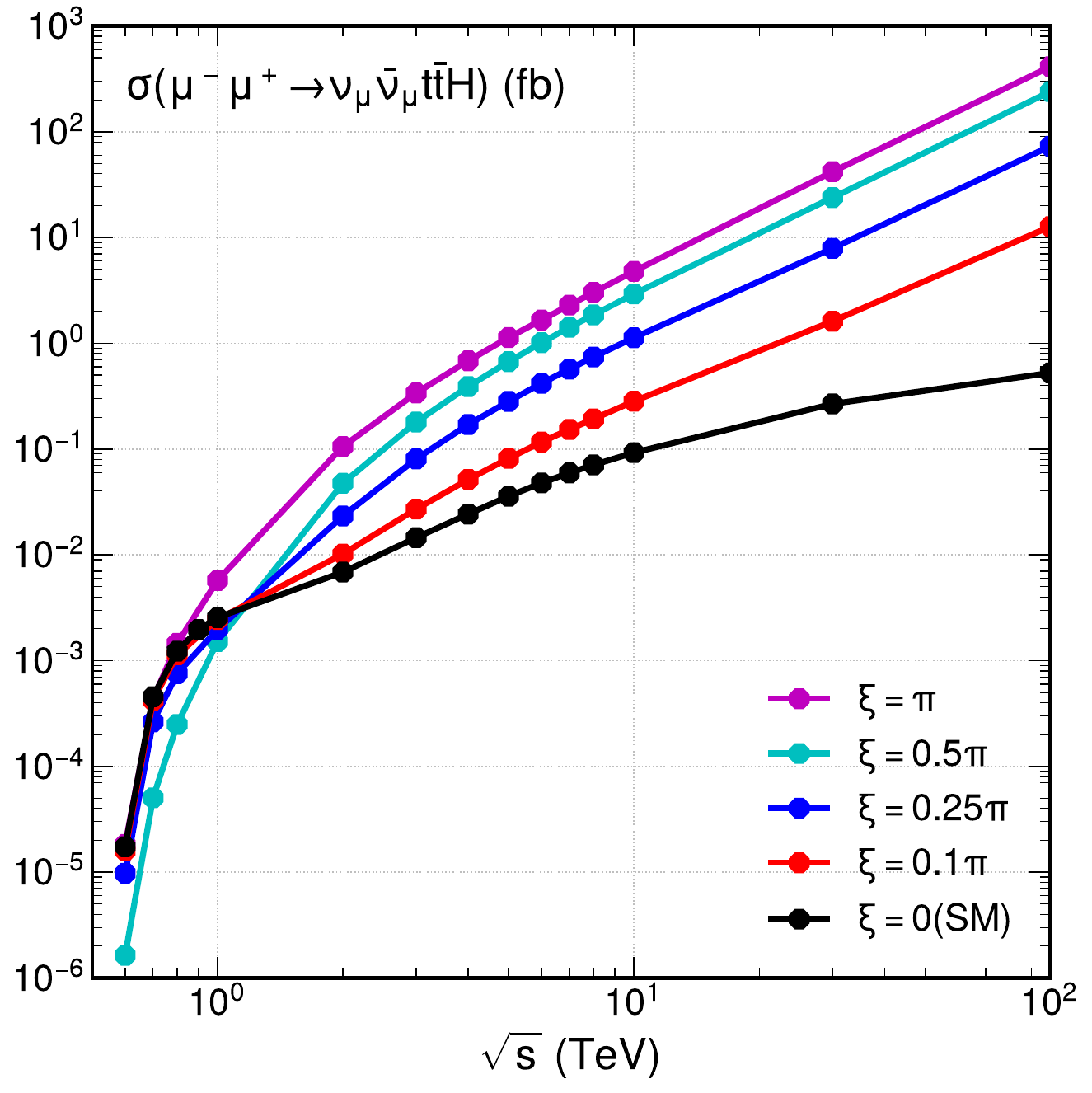}}
\caption{ Same as Fig.\ref{fig:tthvv_CYxs} but for SMEFT.}
\label{fig:tthvv_EFT_xs}
\end{figure}

The total cross section of the muon collider process in Eq.\,(\ref{proc:mmttH})
is now calculated in SMEFT by including the diagram
Fig.\ref{fig:feynman3}(b), which is evaluated numerically by
using the {\tt HELAS} code with the new vertex. 
The results are shown in Fig.\ref{fig:tthvv_EFT_xs}(a) for the $\xi$
dependence at several collision energies, and
in Fig.\ref{fig:tthvv_EFT_xs}(b) for the energy dependence at several
$|\xi|$ values.
When compared with the complex Yukawa model
results of Fig.\ref{fig:tthvv_CYxs}(a) and (b), both the $|\xi|$
dependence and the energy dependence of the
total cross section are milder in SMEFT.
Nevertheless, the strong energy dependence
of the total cross section remains, as a
consequence of the $(\sqrt{\hat{s}})^2$ growth
of the weak boson fusion cross section.
\begin{figure}[b]
\subfigure[]{\includegraphics[width=0.46\textwidth,clip]{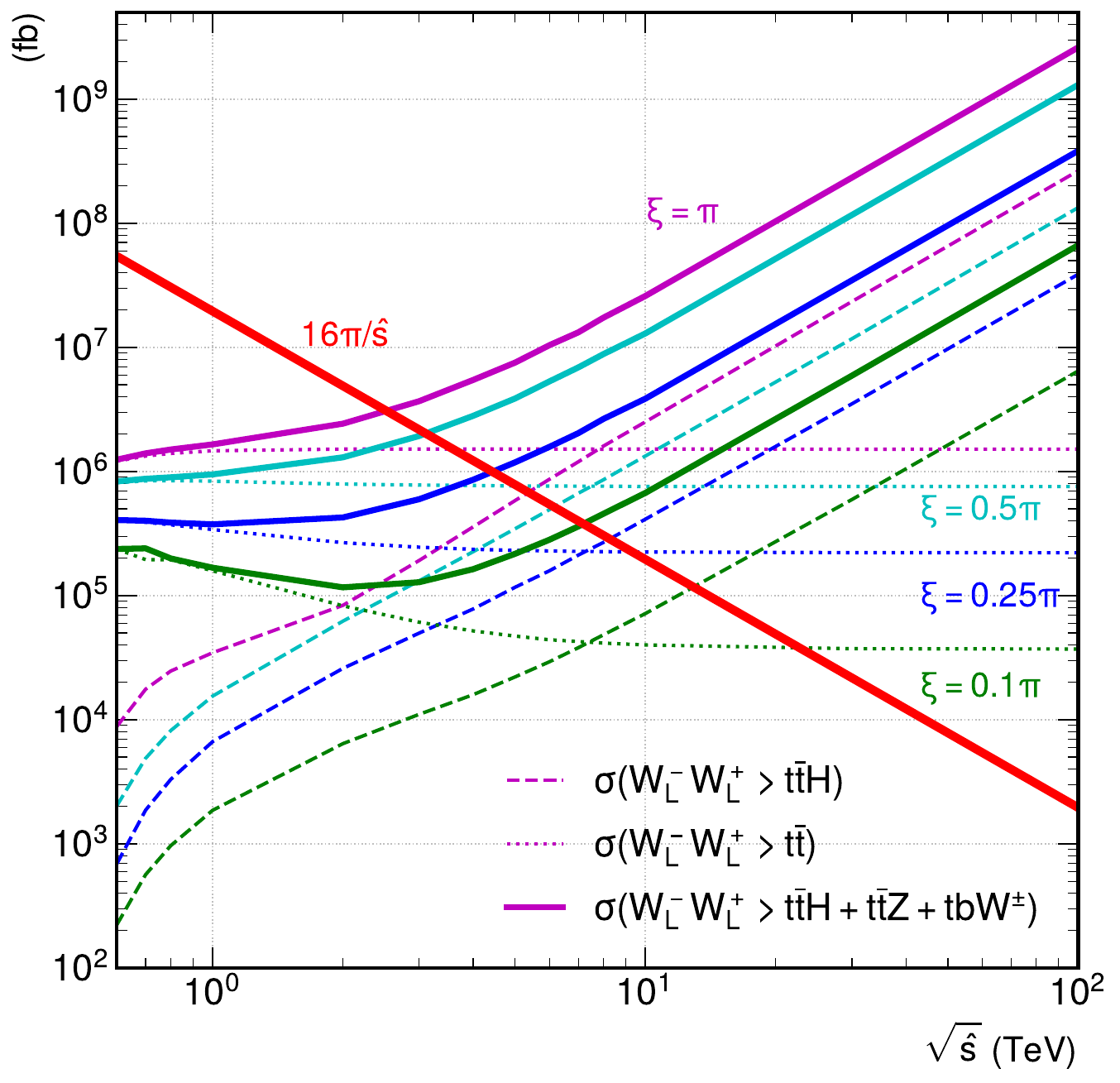}}
\subfigure[]{\includegraphics[width=0.46\textwidth,clip]{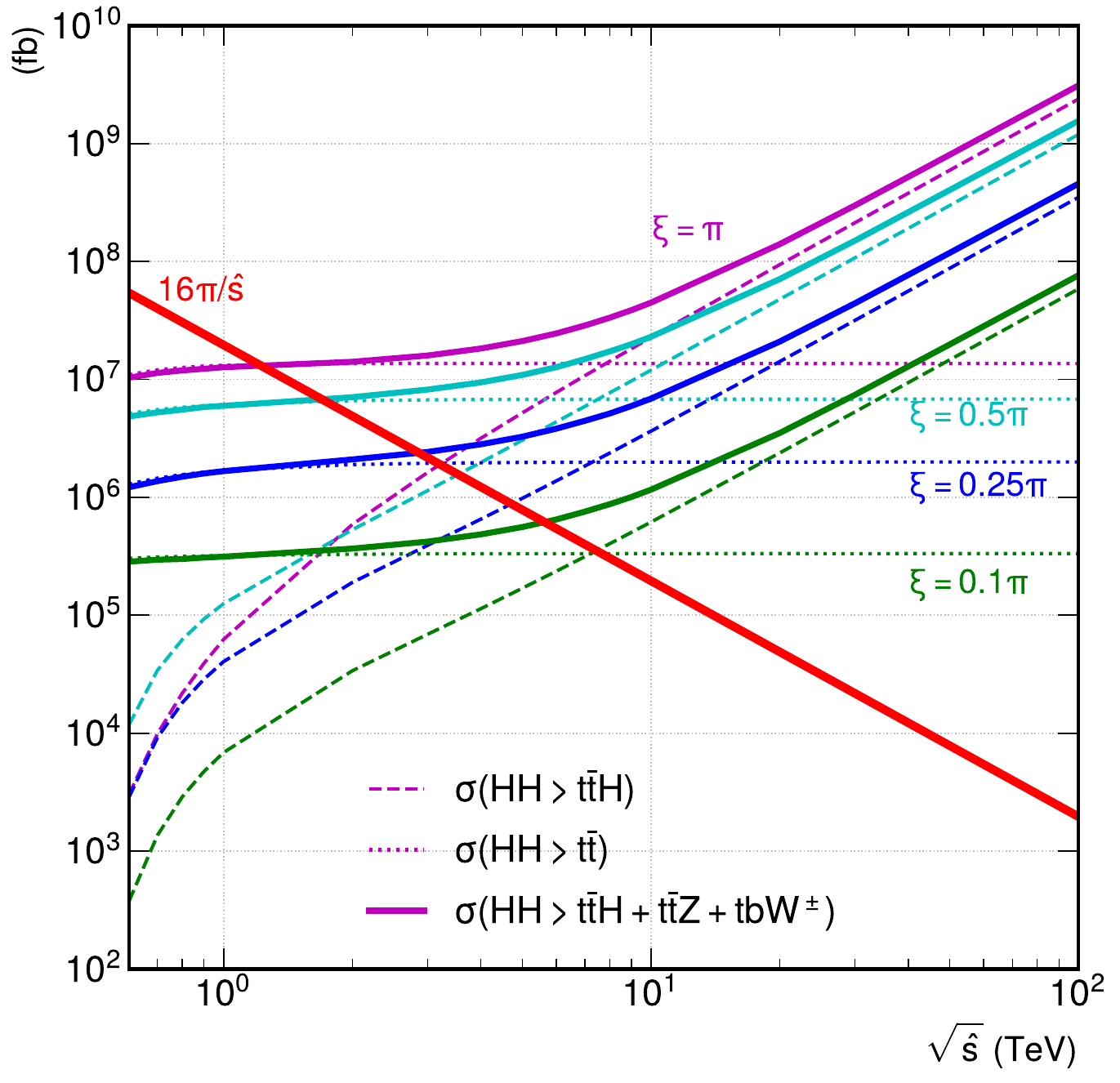}}
\caption{ Perturbative unitarity bound from $\sigma(W_L^-W_L^+\to X)$ and $\sigma(HH\to X)$. }
\label{fig:unitarity}
\end{figure}

Before closing,  we examine perturbative unitarity constraints of the SMEFT model of Eq.\,(\ref{eq:op}).
The high energy amplitudes of the weak boson
fusion process Eq.\,(\ref{proc:wwhtt}) is dominated by the $J=0$
amplitudes, as is clear from the Goldstone
boson amplitudes in Eq.\,(\ref{eq:pipitth}) from the diagram Fig.\,\ref{fig:feynman3}(c).
We can therefore obtain constraints on the
SMEFT operator from the scattering amplitudes
of the $J=0$ state of longitudinally polarized
weak boson pair
$\ket{i} = \ket{ W_L^- W_L^+ (J=0) }$.
In the optical theorem
\begin{eqnarray}
2{\rm Im} \expval{{ T}}{i} = \sum_f \left|\bra{f} { T}\ket{i} \right|^2,
\end{eqnarray}
tells that the final state $\ket{f}$ is summed over
all $J=0$ final states including the phase space
integral.
The unitarity bound
$
\left| {\rm Im} \expval{{ T}}{i}\right| < \left|\expval{{ T}}{i}\right| < 16\pi
$
can then be expressed as
\begin{eqnarray}
\sum_f \sigma_{\rm tot}\left(W_L^- W_L^+ \to f;J=0\right) < \frac{16\pi}{ \hat{s}}.
\label{eq:16pi/s}
\end{eqnarray}
All the SM cross sections fall as $1/\hat{s}$ at high
energies in the $J=0$ channel,
and hence only the contact SMEFT couplings
are relevant.
Examining the SMEFT Lagrangian of Eq.\,(\ref{eq:SMEFTlag}), we find
that only one $2\to2$ process,
$
W_L^- W_L^+ \to t \bar{t}    
$
\cite{Whisnant:1994fh,Chen:2022yiu,Liu:2023yrb},
and
four $2\to3$ processes,
$
W_L^- W_L^+ \to t \bar{t} H,
 t \bar{t} Z, t\bar{b}W^-$ and $\bar{t}bW^+
$,
give non-vanishing total cross section at
high energies.
\begin{table}[t]
\begin{tabular}{|c||c|c|c|c|c|}
\hline
$|\xi|$&$\pi$&$0.5\pi$&$0.25\pi$&$0.1\pi$
\\
\hline
$|\lambda|\cdot \Lambda^{-2}({\rm TeV}^{-2})$&$32.9$&$23.2$&$12.6$ &$5.14$
\\
\hline
\hline
$\sqrt{\hat{s}}_{W_LW_L}({\rm TeV})$&2.5&3.1&$4.4 $&7.2
\\
\hline
$\sqrt{\hat{s}}_{HH}~({\rm TeV})$&1.2&1.7&$2.9$&5.6
\\
\hline
\end{tabular}
\caption{Perturbative unitarity bounds from Fig.\,\ref{fig:unitarity}.}
\label{tab:WWtt_cross_section}
\end{table}

In Fig.\,\ref{fig:unitarity}(a), we show the total cross section of the $t\bar{t}$ and $t\bar{t}H$ production processes in dotted and dashed lines respectively, and the sum
of all the non-decreasing cross sections as a solid line as functions
of colliding $W^-W^+$ energy, $\sqrt{\hat{s}}$, for
several $|\xi|$ values.
The unitarity bound of $16\pi/\hat{s}$ is shown by the
straight solid line in red.
The perturbative unitarity bound of the SMEFT model Eq.\,(\ref{eq:op}) can be read off from the crossing points for each $|\xi|$
value, which are tabulated in Table.\ref{tab:WWtt_cross_section}.
Perturbative unitarity is violated at 2.5 TeV if $\xi=\pi$, or when $|\lambda|/\Lambda^2=32.9$ TeV$^{-2}$, see Eq.\,(\ref{eq:geixi-gsm}).

From the Lagrangian Eq.\,(\ref{eq:SMEFTlag}) the $J=0$ cross sections rise fastest in the $HH$ channel, because of the large $ttHH$ and $ttHHH$ couplings. We show in Fig.\ref{fig:unitarity}(b), the total cross sections of $HH\to t\bar{t}$, $t\bar{t}H$ and all final states in the $J=0$ channel. The perturbative unitarity limit is significantly lower, 1.2 TeV for $|\xi|=\pi$, 5.6 TeV for $\xi=0.1\pi$.
Therefore, when viewing the SMEFT cross sections given in Fig.\ref{fig:WWttH}(b), the curves can be trusted only below the energies in the bottom row of the table.
In the muon collider process Eq.\,(\ref{proc:mmttH}),
we should restrict use of the perturbative
amplitudes to the region where $m(t\bar{t}H)$ is below the bounds in the $WW$ fusion subprocess.

We note that all the SMEFT cross sections presented in this paper have been obtained by treating the Lagrangian Eq.\,(\ref{eq:SMEFTlag}) as a gauge invariant version of the complex Yukawa coupling Eq.\,(\ref{eq:CYlag}). 
When there are two $ttH$ vertices in a Feynman diagram, both are replaced by the complex couplings, and thus the amplitude contains $\Lambda^{-4}$ terms. 
The perturbative unitarity bounds in Table.\ref{tab:WWtt_cross_section} are valid since the $J=0$ amplitudes at high energies are dominated by the dimension five and six vertices. 
Phenomenology of CP asymmetries in the EFT framework are reported elsewhere~\cite{BHKMZ}. 

\section*{Acknowledgement}
We are indebted to the late Cen Zhang for suggesting that we examine the SMEFT operator for the CP violating top Yukawa coupling.
Throughout this study, we appreciated his deep
insights on how we should look for non-Standard
physics.
It is a great loss for us that we cannot learn
from him any more.
We would like to thank Kenichi Hikasa, Kun-Feng Lyu and Kentarou Mawatari for very helpful discussions. 
YJZ thanks Ian Lewis and KC Kong for collaboration at early stage of this work.
VB gratefully acknowledges support from the William F. Vilas Estate.
KH is supported by  the US Japan Cooperation Program in High Energy Physics.
YJZ is supported by JSPS KAKENHI Grant No.\,21H01077 and 23K03403.

\bibliography{tthvv}
\bibliographystyle{utphys}

\end{document}